**Gender differences in scientific careers: A large-scale bibliometric analysis**

Hanjo Boekhout[1], Inge van der Weijden[2], and Ludo Waltman[3]

[1] Leiden Institute of Advanced Computer Science, Leiden University, The Netherlands[*]

h.d.boekhout@liacs.leidenuniv.nl – https://orcid.org/0000-0002-8456-9063

[2] Centre for Science and Technology Studies, Leiden University, The Netherlands

i.c.m.van.der.weijden@cwts.leidenuniv.nl – https://orcid.org/0000-0001-5255-3430

[3] Centre for Science and Technology Studies, Leiden University, The Netherlands

waltmanlr@cwts.leidenuniv.nl – https://orcid.org/0000-0001-8249-1752 (corresponding author)

We present a large-scale bibliometric analysis of gender differences in scientific careers, covering all scientific disciplines and a large number of countries worldwide. We take a longitudinal perspective in which we trace the publication careers of almost six million male and female researchers in the period 1996–2018. Our analysis reveals an increasing trend in the percentage of women starting a career as publishing researcher, from 33% in 2000 to about 40% in recent years. Looking at cohorts of male and female researchers that started their publication career in the same year, we find that women seem to be somewhat less likely to continue their career as publishing researcher than men, but the difference is small. We also observe that men produce on average between 15% and 20% more publications than women. Moreover, in biomedical disciplines, men are about 25% more likely than women to be last author of a publication, suggesting that men tend to have more senior roles than women. Compared with cross-sectional studies, our longitudinal analysis has the advantage of providing a more in-depth understanding of gender imbalances among authors of scientific publications.

---

[*] Most of the work for this paper was performed while the author was affiliated with the Centre for Science and Technology Studies, Leiden University, The Netherlands.

# 1. Introduction

Differences between men and women in their participation in scientific research are extensively being discussed. Underrepresentation of women in the research system may indicate that the system does not sufficiently benefit from the contributions that highly-qualified female researchers could make. To the extent that men and women differ in their research interests, it may also be a signal that the system does not provide sufficient room for the perspectives of women, causing certain topics, ideas, or approaches to be overlooked or marginalized.

In this paper, we present a large-scale analysis of gender differences in scientific careers, covering all scientific disciplines and a large number of countries worldwide. We take a bibliometric approach, studying gender differences through the lens of the publications produced by male and female researchers. Rather than providing a cross-sectional picture of gender differences at a specific point in time, we offer a longitudinal perspective in which we trace the careers of almost six million male and female researchers with at least three publications in the period 1996–2018. Such a longitudinal perspective enables us to analyze time trends in the share of men and women starting a career as publishing researcher. It also allows us to study gender differences in the length of the publication career of a researcher and in the development of publication productivity and seniority in the course of a researcher's career.

There is an extensive bibliometric literature on gender differences in science (for an overview, see Halevi, 2019). While earlier studies usually provided a cross-sectional picture of gender differences (e.g., Filardo et al., 2016; Holman et al., 2018; Jagsi et al., 2006; Larivière et al., 2013; West et al., 2013), more recent studies have started to offer longitudinal perspectives in which the careers of researchers are traced over time. A recent report by Elsevier (2020) for instance analyzed the development of male and female careers that started in 2009, and a recent study by Huang et al. (2020) presented a historical analysis by tracing male and female careers that ended between 1955 and 2010. Huang et al. concluded that gender differences in publication productivity and citation impact can largely be explained by differences in career length.

The importance of longitudinal analyses that account for factors such as career length and publication productivity is illustrated by research into gender differences in self-citations. Based on an analysis of 1.5 million publications that appeared between 1779 and 2011, King et al. (2017) found that men self-cite 56% more than women. However, Mishra et al. (2018) showed that gender differences in self-citations largely disappear when accounting for a researcher's prior number of publications (for additional results pointing in the same direction, see Azoulay & Lynn, 2020), suggesting that men do not disproportionally self-cite compared with women.

Like the work by Huang et al. (2020), Mishra et al. (2018), and others, we offer a longitudinal bibliometric perspective on gender differences in science. By analyzing the publication careers of a large number of researchers in all scientific disciplines, we aim to provide a solid understanding of the relationship between gender and factors such as the length of a researcher's publication career and the publication productivity and seniority of a researcher. Our work distinguishes itself from earlier longitudinal bibliometric studies by taking into account a very large number of researchers (i.e., almost six million male and female researchers) and by focusing on a recent time period (i.e., 1996–2018).

This paper is organized as follows. In Section 2, we provide a detailed discussion of our bibliometric methodology, focusing on the Scopus data that we use, the gender inference approach that we take, and the gender statistics that we calculate. We present the results of our analysis in Section 3. We first report time trends in the share of men and women starting a career as publishing researcher. For cohorts of male and female researchers that started their publication career in the same year, we then show statistics on the length of the career of a researcher and on the development of publication productivity and seniority in the course of a researcher's career. Finally, in Section 4, we discuss the conclusions and limitations of our research.

## 2. Data and methods

In this section, we first discuss the data used in our analysis. We then discuss our approach to gender inference and to the calculation of gender statistics.

### 2.1. Bibliographic database

Our analysis is based on the Scopus database produced by Elsevier. Scopus bulk data is delivered annually by Elsevier to the Centre for Science and Technology Studies (CWTS) at Leiden University. To facilitate large-scale bibliometric analyses, the data is stored in a relational database at CWTS. Our analysis uses Scopus data received in April 2019. The data covers publications from 1996 to early 2019.

Instead of the Scopus database, we also considered using the Web of Science database. However, Scopus has the advantage that it offers more complete data on the first names of the authors of publications. As discussed below, data on first names is essential for gender inference. In Web of Science, data on first names is largely missing for publications that appeared before 2006 (Huang et al., 2020).

If two publications include the same author name, they may have been produced by the same author. However, it is also possible that they have been produced by two different authors that happen to have the same name. The other way around, if two publications do not include the same author name, they probably have not been produced by the same author. However, it may be that the publications have been produced by the same author, but the author may not have reported his or her name in a consistent way. An author name disambiguation algorithm attempts to determine which publications belong to the same author.

Our analysis relies on author identifiers provided by Scopus. These identifiers have been assigned using an author name disambiguation algorithm. In some cases, manual corrections have been made to the identifiers based on feedback from Scopus users. The author name disambiguation algorithm used by Scopus is a black box for which no public documentation is available. Based on an evaluation of 12,000 randomly selected authors, Scopus claims that its author identifiers have a precision of 98.1% and a recall of 94.4% (Baas et al., 2020). In an independent evaluation focused on Japanese authors, a precision and recall of 99.4% and 98.4%, respectively, were obtained (Kawashima & Tomizawa, 2015). For Dutch authors, lower values of 87.3% and 95.9% for precision and recall were found (Reijnhoudt et al., 2014). In a more recent evaluation focused on German authors, a precision and recall of 100.0% and 97.1% were reported (Aman, 2018). In the remainder of this paper, we use the terms 'author' and 'researcher' to refer to authors as defined by author identifiers in Scopus.

### 2.2. Gender inference

For each author in Scopus, we attempted to algorithmically infer a gender. A binary concept of gender was used, distinguishing between men and women. If no gender could be inferred for an author, the gender was considered unknown.

While we believe that algorithmic approaches to gender inference yield valuable insights, we emphasize that these approaches also have important limitations. In particular, algorithms for gender inference reduce gender to a binary concept, they are susceptible to bias, and their transparency is limited. We refer to Lindqvist et al. (2020) for a general discussion of the use of the concept of gender in social science research and to Mihaljević et al. (2019) and Rasmussen et al. (2019) for a discussion of the limitations of algorithmic approaches to gender inference.

Our algorithmic approach to gender inference consists of five steps. We now discuss each of these steps.

*Step 1: Determining country of origin*

In each publication of an author, the author is linked to an affiliation. An affiliation has an address, including a country. If the country that an author is most often associated with in his or her publications is the same as the country that the author is associated with in his or her first publication, we consider this country to be the author's country of origin. Otherwise, we regard the evidence to be insufficient to determine a single country of origin. We then consider all countries that the author is associated with to be countries of origin.

*Step 2: Gender inference based on first name and country of origin*

We clean the first name of an author by removing initials (e.g., 'sonya f.p.' → 'sonya'), separating multiple names (e.g., 'michele luigi' → 'michele' and 'luigi'), and extracting nicknames (e.g., '(joe) yong' → 'joe' and 'yong'). We then provide each of the cleaned names of the author and the author's country of origin as input to three gender inference tools: Gender API (https://gender-api.com/), Gender Guesser (https://pypi.org/project/gender-guesser/), and Genderize.io (https://genderize.io/). In the case of Gender Guesser, input is provided both including (gender_ic) and excluding (gender_ec) the author's country of origin. For a comparison of different gender inference tools, we refer to Santamaría and Mihaljević (2018), who found that especially Gender API has a good performance.

The gender of the author is determined based on the output of the three gender inference tools. This is done according to the rules listed in Table 1. These rules are followed from top to bottom. To make sure that gender inferences are sufficiently accurate, the rules take into account the number of samples based on which a gender inference is made and the probability with which a gender inference is expected to be correct. If the author has multiple cleaned names, we first apply the first rule to each name, then the second rule, and so on, until a gender inference is obtained for one of the names.

Table 1. Rules for gender inference based on first name using Gender API, Gender Guesser, and Genderize.io.

| Rule | Inference |
| --- | --- |
| Gender API: samples > 2, probability ≥ 90%, gender = male/female | male/female |
| Gender API: samples ≤ 10<br>Gender Guesser: gender_ic = male/female | male/female |
| Gender API: samples ≤ 10<br>Gender Guesser: gender_ic = unknown OR no inference, gender_ec = male/female | male/female |
| Gender API: samples ≤ 10<br>Genderize.io: samples > 3, probability ≥ 90%, gender = male/female | male/female |
| Gender API: samples > 0 OR Genderize.io: samples > 0<br>(The gender inference tools yielded results based on too few samples or with a too low probability.) | unknown |
| Otherwise<br>(The gender inference tools yielded no results at all.) | no inference |

If the author has multiple countries of origin, the above procedure is followed separately for each country. The results for the different countries will be combined in step 5 below.

*Step 3: Gender inference based on last name and country of origin*

Larivière et al. (2013) identified eight countries, such as Russia, where last names are indicative of gender. If an author has one of these countries as country of origin and if no gender inference (male or female) has been made for the author in step 2, we use the approach of Larivière et al. to infer a gender for the author based on the author's last name.

*Step 4: Gender inference based on first name and other countries*

If no gender inference (male, female, or unknown) has been made for an author in steps 2 and 3, we look at all authors in the analysis for which a gender inference has been made and that have the same first name but a different country of origin. If the same gender has been inferred for all these authors, the focal author is also considered to have this gender. Otherwise, the gender of the focal author is considered unknown.

As an example, consider an author with first name James and country of origin Taiwan. Suppose that no gender has been inferred for this author in steps 2 and 3. If all other authors with first name James have been inferred to be male, James from Taiwan is also considered to be male.

*Step 5: Combining results for multiple countries of origin*

If an author has multiple countries of origin, the gender inference results for the different countries need to be combined. If the result is the same for each country, this result determines the gender of the author. Otherwise, the gender of the author is considered unknown.

*Evaluation*

To evaluate the accuracy of our approach to gender inference, we used a data set created by Larivière et al. (2013). The results of the evaluation are discussed in Appendix A. Compared with the gender inference approach of Larivière et al., our approach is slightly less accurate for authors inferred to be male and substantially more accurate for authors inferred to be female.

**2.3. Gender statistics**

In the calculation of the gender statistics reported in this paper, we considered only publications of the document types *article*, *chapter*, *conference paper*, *conference review*, and *review* in Scopus. Publications of other document types were disregarded. We considered only authors that have at least three publications. Authors with one or two publications are relatively likely to be the result of mistakes made by Scopus' author name disambiguation algorithm.

For each author, we determined the year of the first publication and the year of the last publication. These years mark the start and the end of the publication career of an author. An important caveat applies. Because publications that appeared before 1996 are not included in our data, the year of someone's first publication could not always be correctly determined. The other way around, because someone may continue to publish in the future, the year of someone's last publication could not always be correctly determined either. For instance, someone's last publication in our data may be from 2017, but this author may have produced another publication in 2020. This author is then incorrectly considered to have ended his or her publication career in 2017. As a consequence of these issues, statistics for the first and last years covered by our analysis are less reliable than statistics for the years in between.

We also linked authors to scientific disciplines. We used the disciplinary classification provided by Scopus for this purpose. In this classification, each source (e.g., a journal or a conference proceedings) in Scopus belongs to one or more disciplines. The classification consists of two hierarchically related levels. We considered only the top level, which comprises 26 disciplines (excluding the category Multidisciplinary). Many authors have publications in different sources that belong to different disciplines. We linked an author to a discipline if at least 80% of the publications of the author have appeared in sources belonging to the discipline. This means that authors with a highly multidisciplinary publication profile were not linked to any discipline. Because a source may belong to more than one discipline, there are also authors that were linked to multiple disciplines. The accuracy of the disciplinary classification of Scopus was criticized by Wang and Waltman (2016). However, because our analysis uses only the top-level disciplines in this classification, we consider the classification to be sufficiently accurate for our analysis. In the presentation of our results, we leave out Decision Sciences because this discipline is much smaller than the other disciplines in the disciplinary classification of Scopus.

Finally, we linked authors to countries. An author was linked to a country if the author has an affiliation with an address in the country in at least 80% of his or her publications. As a result, some authors that have worked in multiple countries were not linked to any country. Because an author may have multiple affiliations in the same publication, possibly with addresses in different countries, there are also authors that were linked to multiple countries.

There are large differences between countries in the share of authors for which the gender is unknown. For countries that have a large share of authors with an unknown gender, we did not calculate gender statistics, since statistics for these countries could be misleading. We decided to include in our analysis only the 95 countries for which the share of authors with an unknown gender does not exceed 33%. Countries such as France and South Korea are just below this threshold and are therefore included in our analysis, while countries such as China and India are above the threshold and are therefore excluded from the analysis. Our general statistics, which do not pertain to a specific country, cover all authors that have in at least 80% of their publications an affiliation with an address in one of the above-mentioned 95 countries.

**3. Results**

We begin this section by analyzing differences between men and women in terms of the start and end of their career as publishing researcher. We then consider gender differences in terms of the publication productivity and seniority of researchers.

Our analysis includes 6.9 million researchers with at least three publications in the period 1996–2018. The results presented in Subsection 3.1 are based on 3.9 million male researchers and 2.1 million female researchers. In most of the results, the 0.9 million researchers for whom the gender is unknown are not included.

In Subsections 3.2, 3.3, and 3.4, we consider only the 205,779, 251,343, and 267,815 male and female researchers that started their career as publishing researcher in, respectively, 2000, 2005, and 2010. By focusing on these three cohorts of researchers, we are able to separate gender differences in recent years from those in earlier years and to account for demographic inertia (Shaw & Stanton, 2012). Statistics for men and women that started their career as publishing researcher in other years can be found in a data set that we have made available (Boekhout et al., 2021).

### 3.1. Career start

Figure 1 shows the time trend in the percentage of men and women starting a career as publishing researcher. The left panel presents the time trend when researchers for whom the gender is unknown are included in the statistics. The right panel presents the time trend when these researchers are excluded. Like the statistics in the right panel, all remaining statistics reported in this paper do not include researchers for whom the gender is unknown.

Figure 1 shows a clear increase in the percentage of women starting a career as publishing researcher. Before 2000, for every two women starting a publication career there were four to five men. In recent years, there were about three men for every two women. The statistics for 1996 and 1997 are not reliable because they include many researchers that started their publication career before 1996. Likewise, the statistics for the most recent years have a low reliability because some researchers that started their publication career in these years did not yet have three publications and are therefore not included in the statistics. Most likely, the drop in the percentage of women in recent years is an artefact resulting from the requirement of having at least three publications, since women tend to produce fewer publications than men (see Subsection 3.3).

Figure 2 presents a breakdown by discipline of the percentage of men and women that started a career as publishing researcher in 2000 or 2010. In all disciplines in the physical sciences, engineering, and mathematics, the number of men that started their publication career in 2010 was much larger than the number of women, ranging from 62% men in Chemistry to 85% men in Engineering. The underrepresentation of women in these disciplines is discussed extensively in the literature, including also in detailed bibliometric studies focused on specific disciplines, such as computer science (Cavero et al., 2015), mathematics (Mihaljević & Santamaría, 2020; Mihaljević-Brandt et al., 2016), and physics and astronomy (Mihaljević & Santamaría, 2020).

As shown in Figure 2, in most disciplines in the biomedical and health sciences, there were no major differences in the number of men and women that started a career as publishing researcher in 2010. In some disciplines, such as Environmental Science and Dentistry, the number of men starting a publication career was somewhat larger than the number of women. In other disciplines, such as Immunology and Microbiology, the situation was the other way around. Nursing was the only exception. In this discipline, women represented 78% of all researchers that started their publication career in 2010.

For disciplines in the social sciences and humanities, Figure 2 offers a more mixed picture. The two extremes are Economics, Econometrics and Finance on the one hand, where women represented only 28% of all researchers that started their publication career in 2010, and Psychology on the other hand, where women represented 66% of all researchers starting a publication career. We refer to Lundberg and Stearns (2019) for an in-depth discussion of the underrepresentation of women in economics. In psychology, the overrepresentation of women among early-career researchers was also observed in a recent analysis by González-Alvarez and Sos-Peña (2020). As can be seen in Figure 2, when looking at researchers that started their publication career in 2010, Arts and Humanities is close to gender parity. However, this may hide significant variation among different Arts and Humanities fields. Women are for instance strongly underrepresented in philosophy (Wilhelm et al., 2018).

For all disciplines except for Nursing, Figure 2 shows that the percentage of women starting a career as publishing researcher was higher in 2010 than in 2000. The difference was largest in Psychology, Environmental Science, and Chemical Engineering. In each of these disciplines, the percentage of women starting a publication career increased by 12 percentage points between 2000 and 2010. Most disciplines moved in the direction of gender parity. However, Psychology is a notable exception (see also González-Alvarez & Sos-Peña, 2020). In this discipline, the percentage of women starting a publication career increased from 54% in 2000 to 66% in 2010.

Figure 3 shows a breakdown by country of the percentage of women that started a career as publishing researcher in 2010. Countries that are colored gray are not covered by our analysis. Percentages below 30% can be observed for some African countries, a number of countries in the Middle East, and also for Asian countries such as Japan (18%) and South Korea (21%). There are 12 countries where more than half of the researchers that started a publication career in 2010 were female. In addition to a number of countries in Eastern and Southern Europe, this was the case for Puerto Rico (51%), Iceland (52%), Tunisia (56%), Philippines (57%), and Argentina (57%). We note that country-level statistics need to be interpreted with some care, because the accuracy of our approach to gender inference may vary between countries. In the remainder of this paper, we do not consider statistics at the level of countries. However, detailed country-level statistics can be found in a data set that we have made available (Boekhout et al., 2021).

**3.2. Career end**

For male and female researchers that started their career as publishing researcher in 2000, 2005, or 2010, Figure 4 shows the annual percentage of researchers ending their career. This percentage was calculated relative to all male or female researchers still active after a certain number of years. For instance, of the 137,715 men that started their publication career in 2000, 120,366 were still active after 5 years and 114,753 were still active after 6 years. Hence, of the men still active after 5 years, (120,366 − 114,753) / 120,366 = 4.7% ended their career in the next year. This was the case for 5.2% of the active women.

For researchers that started their publication career in 2005 or 2010, Figure 4 shows that in the early years of their career men were somewhat more likely to end their career than women. In later years, women were more likely to end their career. For researchers that started their publication career in 2000, women consistently had a higher probability to end their career than men (except for Y = 1 in Figure 4), but in later years of their career the difference was very small. At the end of all three time trends in Figure 4, there is a substantial increase in the percentage of men and women ending their publication career. This seems to be an artefact resulting from researchers that are incorrectly considered to have ended their career in recent years. These researchers have no publications in the most recent years covered by our analysis, but they will have publications in future years not covered by our analysis. Figure 4 also shows that the probability of a researcher ending his or her publication career is lowest for the oldest cohort of researchers and highest for the youngest cohort. This may reflect the rise of the temporary workforce in science reported by Milojević et al. (2018).

Figure 5 presents a breakdown by discipline of the percentage of men and women ending their career as publishing researcher within five years after they started their career. Statistics are reported for two cohorts of researchers, a cohort of researchers that started their career in 2000 and a cohort of researchers that started their career 2010. We focus on a period of five years instead of a longer period because this enables us to compare these two cohorts of researchers. Similar choices are made in Figures 7, 8, 10, and 11 discussed in the next subsections. Statistics for longer periods can be found in a data set that we have made available (Boekhout et al., 2021).

As can be seen in Figure 5, for researchers starting their career in 2010, gender differences were small in most disciplines. In some disciplines men were somewhat more likely to end their career, while in other disciplines women had a somewhat higher probability to end their career. The difference between men and women was more substantial in Nursing (44% vs. 35%) and Business, Management and Accounting (27% vs. 21%), with men being more likely to end their career than women in both disciplines. For researchers that started their publication career in 2000, the situation was less balanced. As can be seen in Figure 5, in 19 of the 25 disciplines men were less likely to end their career than women. The difference was largest in Psychology (13% vs. 20%) and Engineering (26% vs. 32%).

Results similar to ours were reported by Elsevier (2020) for researchers that started their publication career in 2009. Huang et al. (2020) also performed a similar analysis, but their findings are different from ours. They found that each year women are almost 20% more likely to end their publication career than men. The analysis of Huang et al. covers an older time period, which may explain why their findings are different. For assistant professors hired since 1990 at US universities, it was found that both in science and engineering (Kaminski & Geisler, 2012) and in the social sciences (Box-Steffensmeier et al., 2015) there are hardly any gender differences in retention rates. This seems to align with our finding that in later years of their career (i.e., after the PhD and postdoctoral period) female researchers that started their career in 2000 were only slightly more likely to end their career than their male colleagues.

### 3.3. Publication productivity

For men and women that started their career as publishing researcher in 2000, 2005, or 2010, Figure 6 shows the time trend in their average annual number of publications. For all researchers that were still active in a certain year of their career, the figure presents the average number of publications they produced in that year. The left and the right panel present statistics based on, respectively, a full counting and a fractional counting approach. To illustrate the difference between these two counting approaches, we consider a publication that is co-authored by *n* researchers. In the full counting approach, this publication is counted fully for each of the *n* researchers. Each researcher is considered to have produced a full publication. In contrast, in the fractional counting approach, the publication is counted fractionally for each of the *n* researchers, and each researcher is considered to have produced (1 / *n*)th of a full publication.

Figure 6 shows that men consistently had a substantially higher publication productivity than women, regardless of the year in which they started their career as publishing researcher and regardless of the period in their career. Based on the full counting statistics, the publication productivity of men was between 20% and 35% higher than the publication productivity of women. Based on the fractional counting statistics, the difference was even larger, between 25% and 40%.

There are well-known differences between disciplines in the average number of publications of a researcher. The gender differences observed in Figure 6 could be due to a relative overrepresentation of men in disciplines with a larger average number of publications per researcher. Figures 7 and 8 can be used to explore this possibility. These figures present a breakdown by discipline of the average number of publications of men and women in year 6 of their publication career. Statistics are reported for researchers that started their publication career in 2000 or 2010, based on both a full counting (Figure 7) and a fractional counting (Figure 8) approach.

We first consider the full counting statistics in Figure 7. In 22 of the 25 disciplines, male researchers that started their publication career in 2000 had a higher publication productivity in year 6 of their career than their female colleagues. A similar observation can be made for researchers that started their publication career in 2010. On average, the publication productivity per discipline was 17% higher for male researchers that started their career in 2010 than for their female colleagues (calculated by averaging the percentage difference over all disciplines, weighing each discipline by its number of researchers). Hence, the overall gender difference in publication productivity observed in the left panel of Figure 6 can also be found at the level of individual disciplines. However, at the disciplinary level, differences between men and women were somewhat smaller than at the overall level. This means that the overall gender difference in publication productivity was caused partly by a relative overrepresentation of men in disciplines with a larger average number of publications per researcher. In particular, the average number of publications per researcher was much larger in Physics and Astronomy than in other disciplines (see Figure 7), and Physics and Astronomy was also one of the disciplines with the largest overrepresentation of men (see Figure 2).

Like the full counting statistics in Figure 7, the fractional counting statistics in Figure 8 indicate that in most disciplines men produced more publications than women in year 6 of their career as publishing researcher. The fractional counting approach yields somewhat larger gender differences than the full counting approach. For researchers that started their publication career in 2010, the publication productivity per discipline was on average 20% higher for men than for women. Based on the fractional counting approach, the average number of publications per researcher was substantially larger in disciplines in the physical sciences, engineering, and mathematics than in disciplines in the biomedical and health sciences. The overrepresentation of men was also much larger in the physical sciences, engineering, and mathematics than in the biomedical and health sciences (see Figure 2). This means that the overall gender difference in publication productivity observed in the right panel of Figure 6 was caused partly by gender differences in productivity at the level of individual disciplines and partly by the relative overrepresentation of men in certain disciplines, in particular in the physical sciences, engineering, and mathematics.

Differences between men and women in publication productivity may at least partly be due to women having more gaps in their publication career than men, for instance because of pregnancy and maternity leave. We analyze this possibility in Appendix B, where we indeed find that women had more gaps in their publication career than men. However, the differences were relatively small and therefore offer only a partial explanation for differences in publication productivity.

Gender differences in publication productivity, with men on average producing more publications than women, have been found in a large number of studies (for an overview, see Halevi, 2019). These differences, which lead to the so-called productivity puzzle (Cole & Zuckerman, 1984), were also observed in recent large-scale analyses by Elsevier (2020) and Mishra et al. (2018). In line with our results reported above, the literature generally finds that gender differences in publication productivity cannot be explained by differences in career length or career stage. However, a notable exception is a recent study by Huang et al. (2020). Based on a large-scale longitudinal analysis, Huang et al. found that differences in career length explain differences between men and women in publication productivity, leading to the conclusion that "men and women publish at a comparable annual rate", referred to as a 'gender invariant' by Huang et al. These findings are in stark contrast with our findings presented above as well as the findings of many other studies. The discrepancy may be due to the older time period covered by the analysis of Huang et al. Moreover, unlike our analysis, the analysis of Huang et al. did not consider the publication productivity of researchers in a specific period in their career, but only their average annual productivity during their entire career. This may also explain why the findings of Huang et al. are different.

**3.4. Seniority**

To analyze gender differences in the seniority of researchers, we use a researcher's position in the list of authors of a publication as a crude proxy of seniority. We distinguish between being the only author of a publication ('single authorship'), being the first author or the last author, or being one of the middle authors. Our focus is on first and last authorship. In most disciplines, being first author of a publication indicates that one has made a major contribution to a research project, either in a more junior or in a more senior role. Being last author of a publication is especially relevant in biomedical disciplines, where the last author typically is a senior researcher that carries the overall responsibility for a research project. Although statistics on first and last authorship provide useful information, we emphasize that there are important differences between disciplines in the way in which the order of the authors of a publication is determined. In some disciplines, in particular in economics, high energy physics, and mathematics, it is quite common to list authors in alphabetical order (Waltman, 2012). First and last authorship have little or no meaning in these disciplines. We refer to Marušić et al. (2011) for an overview of the literature on authorship practices.

Earlier large-scale studies of gender differences in first and last authorship were performed by Holman et al. (2018) and West et al. (2013). Studies with a specific focus on medical journals were carried out by Filardo et al. (2016) and Jagsi et al. (2006), among others. Importantly, unlike our analysis reported below, earlier studies were of a cross-sectional nature. These studies therefore could not account for factors such as the career length of researchers. An alternative approach to study gender differences in seniority is to analyze contributorship (Larivière et al., 2021; Macaluso et al., 2016; Robinson-Garcia et al., 2020). However, given the limited availability of data on contributorship, we do not take this approach.

For men and women that started their career as publishing researcher in 2000, 2005, or 2010, Figure 9 shows the time trend in the probability of being first (left panel) or last (right panel) author of a publication. In the early years of their career, researchers were relatively likely to be first author of a publication. In later years, the probability of being first author decreased and researchers were more likely to be last author. However, there were significant gender differences. As can be seen in the left panel of Figure 9, in the early years of their career, men were more likely than women to be first author of a publication. The opposite pattern can be observed for later years. Moreover, as shown in the right panel of Figure 9, both in earlier and in later years of their career, men were substantially more likely to be last author than women. The probability of being last author was between 15% and 30% higher for men than for women. Since last authorship is an indication of seniority, at least in biomedical disciplines, this suggests that on average men reached senior roles more quickly than women.

As already mentioned, different disciplines have different norms for determining the order of the authors of a publication. For men and women in year 6 of their career, Figure 10 presents a breakdown by discipline of the probability of being first author of a publication. Figure 11 shows the results for last authorship. The figures report statistics for researchers that started their publication career in 2000 or 2010.

From the viewpoint of understanding gender differences in researchers' seniority, the most interesting results are the statistics in Figure 11 for last authorship in the biomedical and health sciences. Looking at researchers that started their publication career in 2010, we observe that in all disciplines in the biomedical and health sciences men had a higher probability than women to be last author of a publication in year 6 of their career. For researchers that started their publication career in 2000, this was the case in 9 of the 11 disciplines in the biomedical and health sciences. Medicine is by far the largest discipline in the biomedical and health sciences, in terms of the number of researchers working in the discipline, followed by Biochemistry, Genetics and Molecular Biology and Agricultural and Biological Sciences. In each of these disciplines, for researchers that started their publication career in 2010 and that were in year 6 of their career, the probability to be last author of a publication was about 25% higher for men than for women, suggesting that on average men moved to senior roles more quickly than women. Underrepresentation of women among last authors was also observed by Holman et al. (2018) and West et al. (2013), but these cross-sectional studies did not account for differences in career length and other factors.

## 4. Discussion

We end this paper by summarizing our main findings, by discussing our conclusions, and by reflecting on the limitations of our research.

### 4.1. Summary

The large-scale bibliometric analysis presented in this paper contributes to a better understanding of gender differences in scientific careers, in particular in the careers of men and women as publishing researchers. We first considered time trends in the share of men and women starting a career as publishing researcher. For cohorts of male and female researchers that started their publication career in the same year, we then examined gender differences in the length of the career of a researcher and in the development of publication productivity and seniority in the course of a researcher's career.

The percentage of women starting a career as publishing researcher shows a clear increasing trend. About 40% of all researchers that started their publication career in recent years were female. In 2000 this was the case for only 33% of the researchers, and before 2000 the percentage was even lower. There are large differences between disciplines. In the physical sciences, engineering, and mathematics, only 22% of the researchers that started their publication career in 2010 were female. In economics, this was the case for just 28% of the researchers. In contrast, in nursing and psychology, women represented, respectively, 78% and 66% of the researchers that started their publication career in 2010. Differences between countries are large as well. In some countries, for instance in Eastern and Southern Europe, more than half of the researchers that started their publication career in 2010 were female. In other countries, such as Japan and South Korea, this was the case for only one out of five researchers.

Women seem to be somewhat less likely to continue their career as publishing researcher than men, but the difference is small. The difference is most clearly visible for researchers that started their publication career in 2000. For researchers that started in 2005 or 2010, the difference is less clear. In the first few years of the careers of these researchers, women were actually more likely to remain active as publishing researcher than men. In later years, however, women had a lower probability to continue their publication career than men. Even for researchers that started their publication career in 2000, the gender difference is limited. As mentioned above, 33% of the researchers that started in 2000 were female. After 15 years, 52% of the male researchers and 54% of the female researchers had discontinued their publication career, resulting in 32% of the researchers still active as publishing researcher being female.

There are substantial gender differences in publication productivity, defined as the number of publications produced by a researcher in a certain year of his or her career. Depending on the year in which they started their career as publishing researcher, the period of their career, and the counting approach (i.e., full or fractional counting), men produced on average between 20% and 40% more publications than women. This can partly be explained by the overrepresentation of men in disciplines with a larger average number of publications per researcher, in particular in the physical sciences, engineering, and mathematics. After correcting for this, we found that in year 6 of their publication career men produced on average between 15% and 20% more publications than women.

Determining the seniority of a researcher is challenging in a bibliometric analysis, but being last author of a publication can be used as a crude proxy of having a senior role, at least in biomedical disciplines. Gender differences in last authorship are substantial. Overall, men were between 15% and 30% more likely to be last author of a publication than women. For researchers in biomedical disciplines that started their publication career in 2010 and that were in year 6 of their career, the probability of being last author was about 25% higher for men than for women. These statistics suggest that men moved to senior roles more quickly than women.

**4.2. Conclusions**

By analyzing the scientific publications produced by male and female researchers, bibliometric studies provide an incomplete but nevertheless very relevant picture of gender differences in science. Many bibliometric studies (for an overview, see Halevi, 2019) have found gender imbalances among authors of publications. However, due to the cross-sectional nature of most of these studies, it is usually difficult to explain these gender imbalances in terms of underlying factors, such as differences in the number of men and women starting a career as publishing researcher or differences in the publication productivity of men and women. Our longitudinal analysis enables us to provide some insight into these factors.

Looking at the various factors considered in our analysis, the overall picture that emerges is that differences in the number of men and women starting a career as publishing researcher are the most important explanation of gender imbalances among authors of scientific publications. In recent years, the number of male researchers starting a publication career was between 40% and 50% larger than the number of female researchers. In earlier years, the difference was even larger, and the effect of this is still visible in the current composition of the workforce of publishing researchers. It will be interesting to see whether the proportion of women starting a career as publishing researcher will continue to increase in the coming years. We also found that men produce on average between 15% and 20% more publications than women. This is another important explanation of gender imbalances among authors of scientific publications. Gender differences in the probability of continuing a career as publishing researcher are small. These differences do not play a significant role in explaining gender imbalances in the scientific literature. Finally, gender differences in last authorship of publications seem to reflect significant differences in the career trajectories followed by male and female researchers, with male researchers being more likely to move to senior roles than their female colleagues.

An approach similar to ours was taken in a recent study by Huang et al. (2020). Like the work reported in the current paper, Huang et al. presented a large-scale bibliometric analysis offering a longitudinal perspective on gender differences in scientific careers. Importantly, however, their findings are markedly different from ours. Huang et al. found that each year female researchers are almost 20% more likely to discontinue their publication career than their male colleagues. This difference is much larger than the difference identified in our analysis. Moreover, while we found that in each year of their career as publishing researcher men on average produce substantially more publications than women, Huang et al. instead referred to publication productivity as a 'gender invariant'. They observed that men tend to produce more publications than women in the course of their entire career as publishing researcher, but they also claimed that, after correcting for differences in career length, "men and women publish at a comparable annual rate". Importantly, Huang et al. considered the average annual publication productivity of male and female researchers during their entire publication career. We instead compared the publication productivity of men and women by looking at specific years in their career as publishing researcher, which yields a more fine-grained perspective on gender differences in productivity. Another important difference is that the analysis of Huang et al. covers an older time period, while our analysis focuses on a more recent period.

The factors identified in our analysis to explain gender imbalances among authors of scientific publications of course also need to be explained themselves. For instance, why are there more men than women starting a career as publishing researcher? And why do men on average produce more publications than women? Questions like these cannot be answered using a bibliometric approach, and we therefore do not address them in this paper. They have been discussed extensively elsewhere (e.g., Ceci et al., 2014; Cheryan et al., 2017; Wang & Degol, 2016).

### 4.3. Limitations

It is essential to be aware of the limitations of our research. First of all, our work is subject to the limitations of the Scopus database. The coverage of the scientific literature provided by the Scopus database is incomplete (Visser et al., 2021). Some researchers therefore are not included in our analysis, while for other researchers their publication oeuvre is covered only partly. In addition, the Scopus data to which we have access does not go back before 1996. This means that our analysis does not properly cover researchers with long careers in science. The disciplinary classification provided by Scopus also has weaknesses, such as questionable classifications of some journals (Wang & Waltman, 2016) and large size differences between disciplines.

Our analysis relies on algorithmic approaches to author name disambiguation and gender inference. These approaches inevitably make mistakes that affect our analysis. For instance, while Scopus' author name disambiguation algorithm appears to be quite accurate (e.g., Baas et al., 2020), the algorithm may sometimes fail to recognize that two authors in fact represent the same person. To reduce the effect of such mistakes, authors with only one or two publications are not included in our analysis, which is another limitation of our work. Our approach to gender inference has the important limitation of relying on a binary notion of gender (Mihaljević et al., 2019; Rasmussen et al., 2019). Moreover, for many authors our approach is unable to infer a gender, and for a substantial number of countries, including large ones such as China and India, we have no gender statistics at all.

While a large-scale bibliometric analysis like the one presented in this paper offers valuable high-level insights into gender differences in scientific careers, it does not provide a more in-depth understanding of gender differences in specific disciplines, research areas, countries, organizations, etc. To obtain such an understanding, our work needs to be complemented with case studies that zoom in on specific populations of researchers. Also, our bibliometric analysis does not take into account the diversity in scientific careers, with different researchers specializing in different roles, which naturally lead to different career trajectories (Robinson-Garcia et al., 2020).

Finally, we emphasize that the bibliometric perspective provided in this paper considers only the scientific publications produced by researchers and does not take into account other activities and outputs, for instance activities related to teaching, consultancy, public outreach, and management and outputs such as data sets, software, lectures, and policy reports. While scientific publications are often seen as the primary output of scientific research, it is essential to be aware that the focus of our analysis on scientific publications yields an incomplete picture of the activities and outputs of researchers and of their career development.

**Acknowledgments**

We are grateful to colleagues at CWTS, Leiden University for their feedback on the research presented in this paper. We thank Vincent Larivière for sharing the validation data set used in Appendix A.

**Author contributions**

Hanjo Boekhout: conceptualization, data curation, formal analysis, investigation, methodology, software, visualization, writing – review & editing

Inge van der Weijden: conceptualization, methodology, writing – review & editing

Ludo Waltman: conceptualization, methodology, visualization, writing – original draft

**Competing interests**

We have no competing interest.

**Funding information**

No funding has been received for the research presented in this paper.

**Data availability**

The research presented in this paper uses Scopus data made available by Elsevier to CWTS, Leiden University. We are not allowed to share this data. The statistics reported in this paper, as well as additional statistics not discussed in this paper, have been made available in a data repository (Boekhout et al., 2021).

**Appendix A: Evaluation of our approach to gender inference**

To evaluate our algorithmic approach to gender inference described in Subsection 2.2, we used a validation data set created by Larivière et al. (2013). Of the 3,704 authors in this data set, 2,351 were included in our analysis. For each of these 2,351 authors, a gender inference (male, female, or unknown) has been made.

Table A1 reports the number and the percentage of correct and incorrect gender inferences. Our gender inference approach has provided an incorrect result for 2.4% of the authors inferred to be male and for 4.2% of the authors inferred to be female. For comparison, Table A1 also shows the results of the gender inference approach of Larivière et al. for the same set of 2,351 authors. The approach of Larivière et al. has provided an incorrect result for 1.8% of the authors inferred to be male, which means that for authors inferred to be male their approach is slightly more accurate than ours. However, for authors inferred to be female, the approach of Larivière et al. has provided an incorrect result in 12.3% of the cases, yielding an accuracy that is substantially lower than ours.

Table A1. Algorithmically inferred gender of an author (in the rows; male, female, or unknown) vs. gender of an author in the validation data set of Larivière et al. (2013) (in the columns; male or female). Statistics for our approach to gender inference and for the approach of Larivière et al.

|         | Our approach | | | | Approach of Larivière et al. (2013) | | | |
|---------|------|--------|------|--------|------|--------|------|--------|
|         | **Male** | | **Female** | | **Male** | | **Female** | |
| **Male**    | 986 | (97.6%) | 24  | (2.4%)  | 533 | (98.2%) | 10  | (1.8%)  |
| **Female**  | 30  | (4.2%)  | 684 | (95.8%) | 85  | (12.3%) | 604 | (87.7%) |
| **Unknown** | 422 | (67.3%) | 205 | (32.7%) | 820 | (73.3%) | 299 | (26.7%) |

**Appendix B: Career gaps**

In this appendix, we present an analysis of gaps in the publication careers of male and female researchers. We define a gap as a period of one or more consecutive years in the publication career of a researcher in which the researcher produced no publications. Women may potentially have more gaps in their publication career than men, for instance because of pregnancy and maternity leave.

To analyze gender differences in gaps in the publication careers of researchers, we considered cohorts of male and female researchers that produced their first publication in 2000, 2005, or 2010 and that had a publication career of at least 5, 10, or 15 years. For each of these cohorts, we calculated the average number of gaps in the first 5, 10, or 15 years of their publication career, distinguishing between gaps of different length (i.e., one year, two years, etc.). Figure B1 presents the results of this analysis.

The results consistently show that publication gaps were more common for female researchers than for their male colleagues, but the differences are relatively small. The number of one-year gaps was between 3% and 6% larger for women than for men. For two-year gaps, the difference was about 10%.

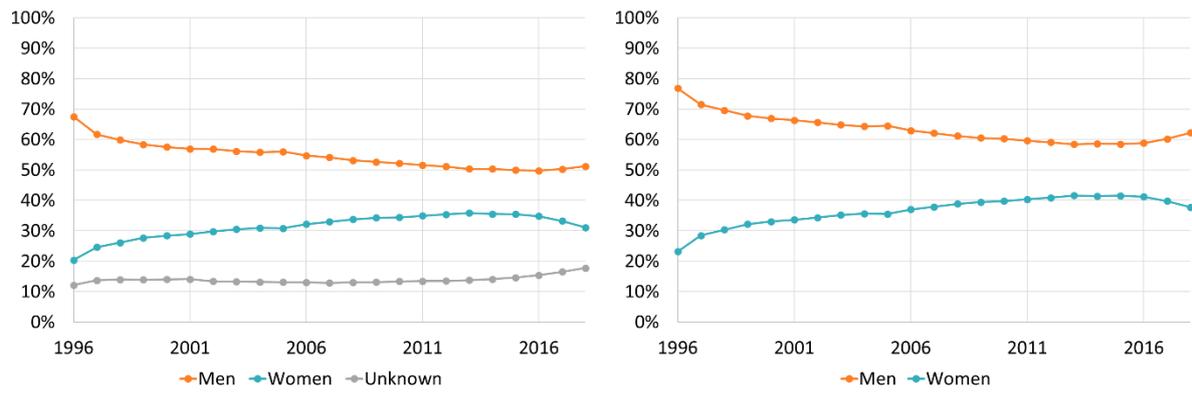
Figure 1. Percentage of men and women starting a career as publishing researcher. Annual statistics for the period 1996–2018, including (left) or excluding (right) researchers for whom the gender is unknown.

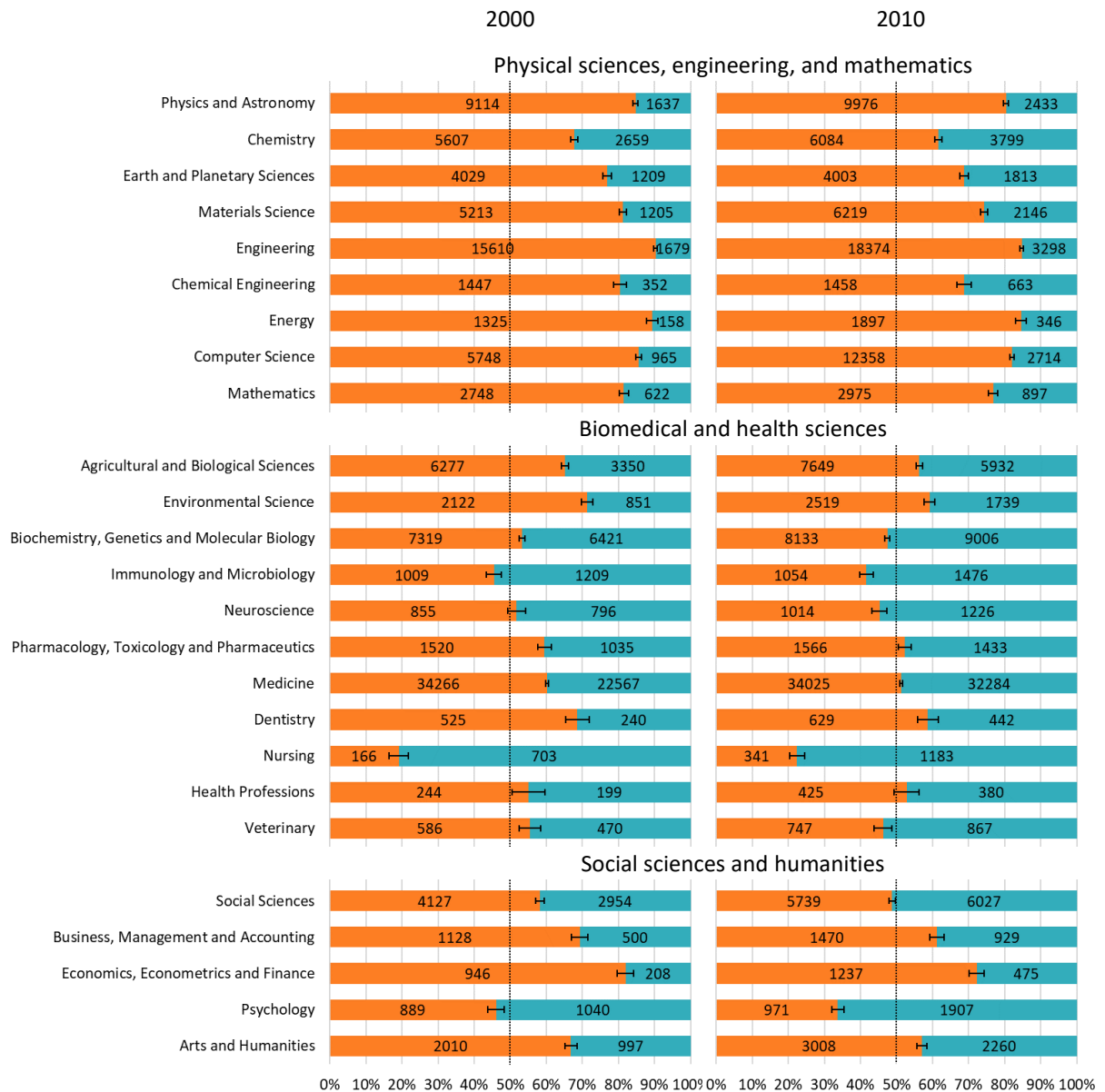

Figure 2. Percentage of men and women starting a career as publishing researcher. Disciplinary statistics for researchers that started their publication career in 2000 or 2010. Error bars represent 95% confidence intervals.

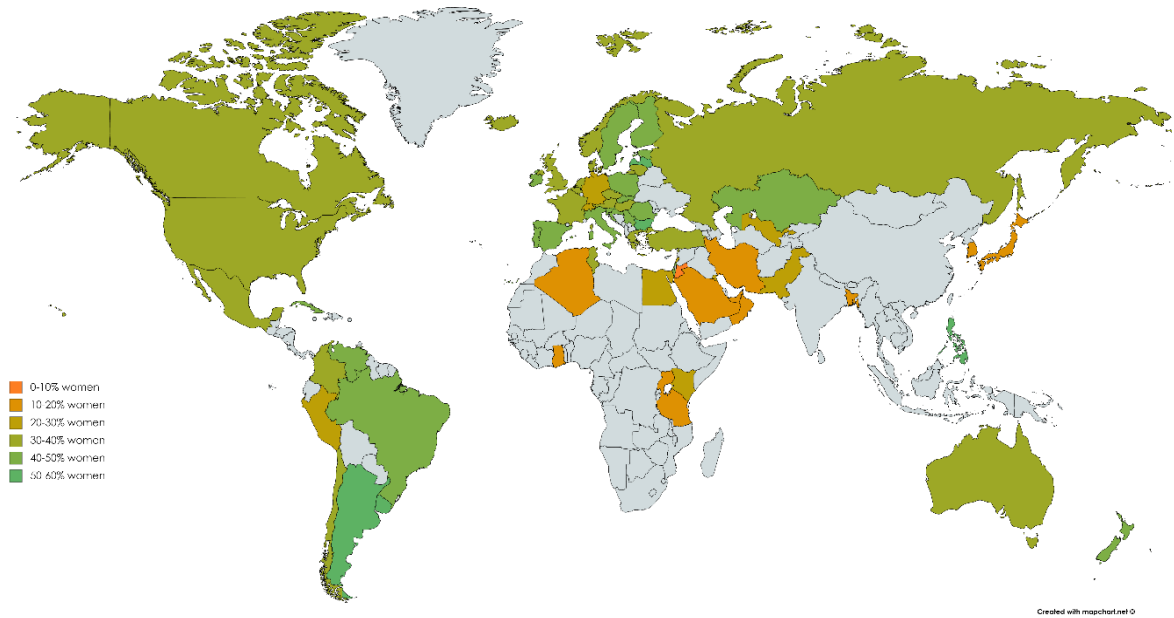

Figure 3. Percentage of women starting a career as publishing researcher. Country statistics for researchers that started their publication career in 2010.

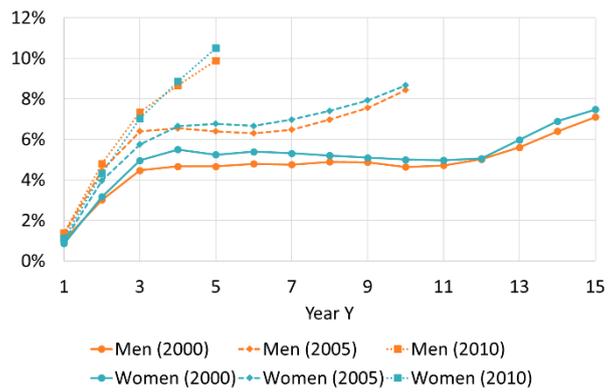

Figure 4. Percentage of men and women that ended their career as publishing researcher after Y years relative to all men or women that were still active after Y years. Statistics for researchers that started their publication career in 2000, 2005, or 2010.

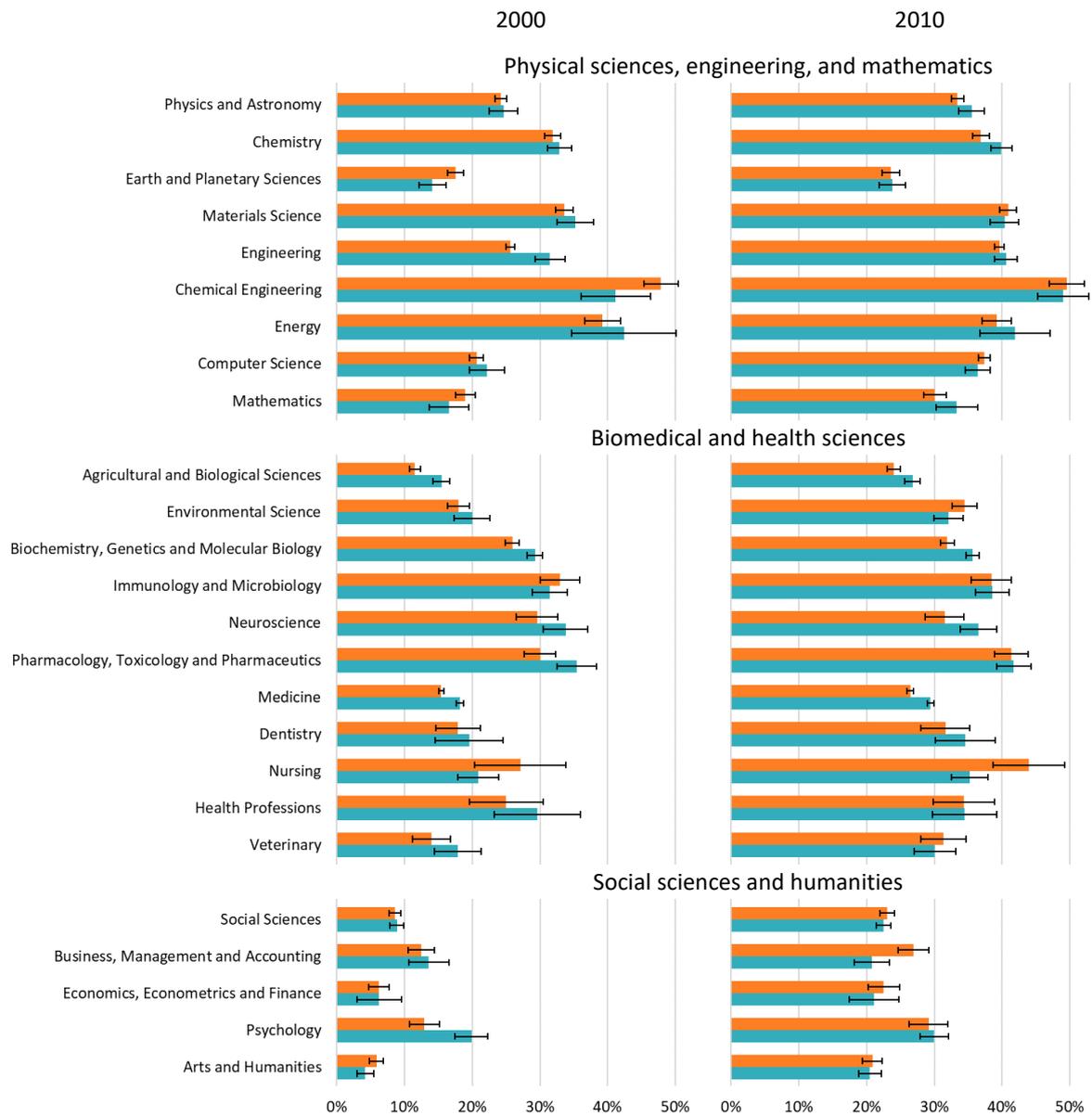

Figure 5. Percentage of men and women that ended their career as publishing researcher within five years after they started. Disciplinary statistics for researchers that started their publication career in 2000 or 2010. Error bars represent 95% confidence intervals.

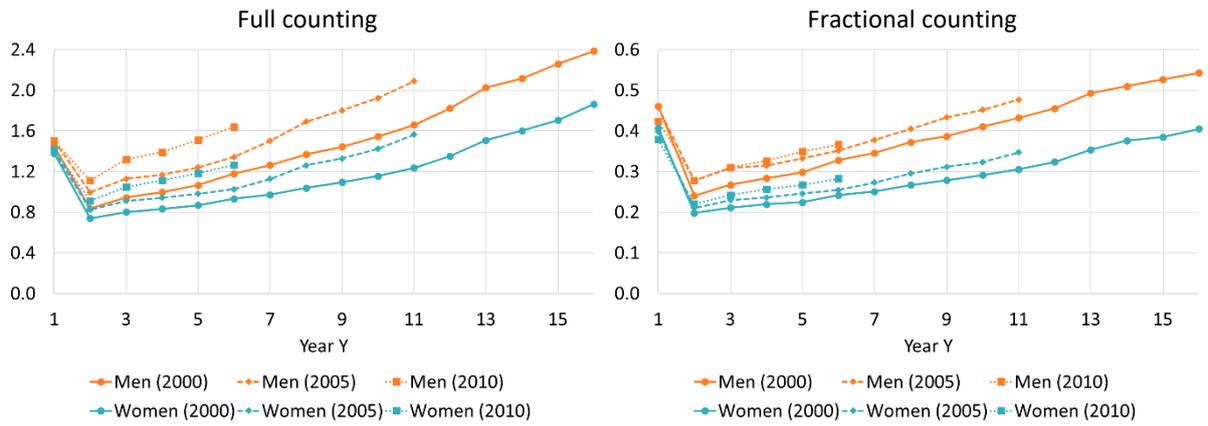

Figure 6. Average number of publications of men and women in year Y of their career as publishing researcher. Statistics for researchers that started their publication career in 2000, 2005, or 2010 based on a full counting (left) or a fractional counting (right) approach.

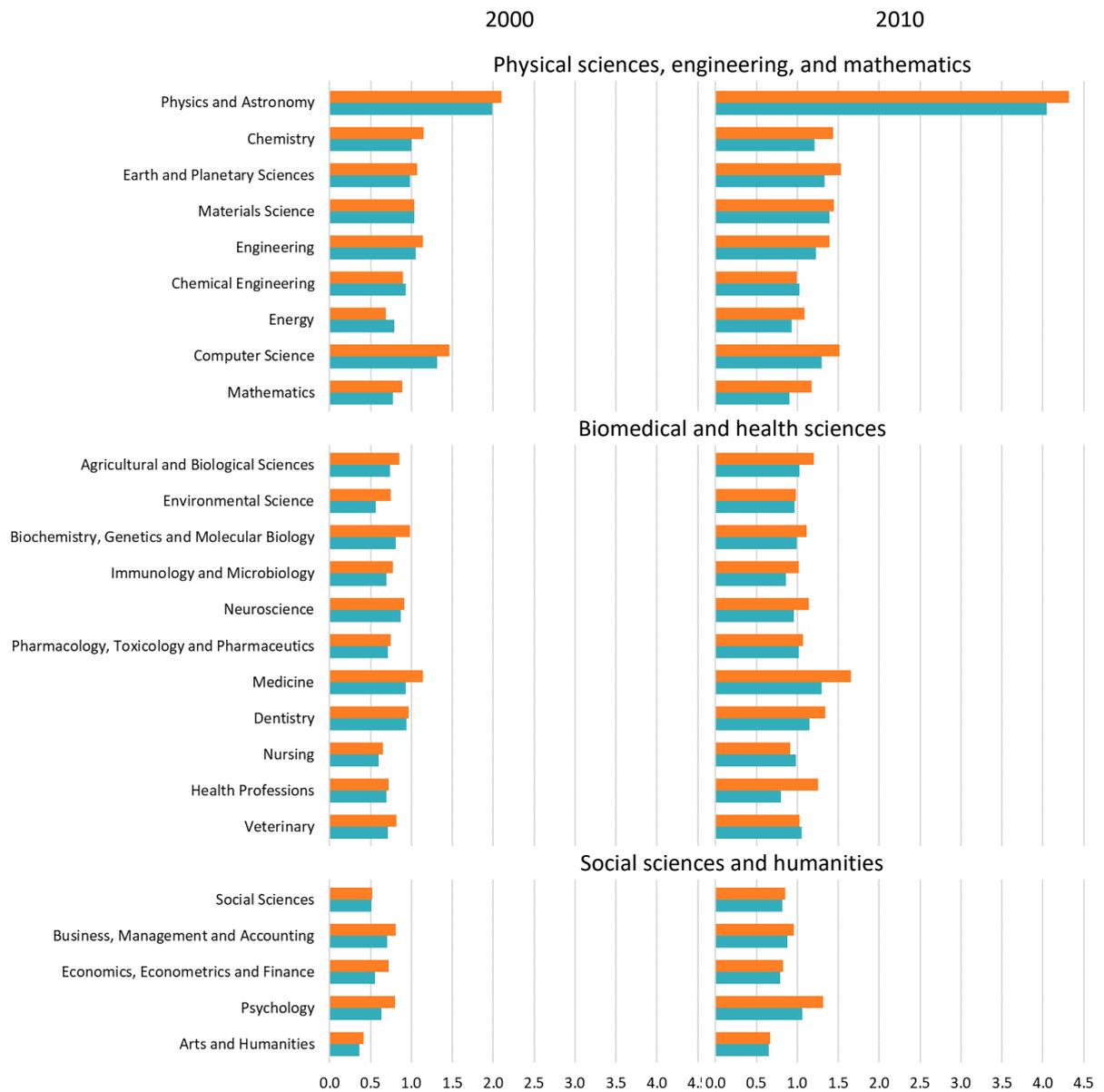

Figure 7. Average number of publications of men and women in year 6 of their career as publishing researcher. Disciplinary statistics for researchers that started their publication career in 2000 or 2010 based on a full counting approach.

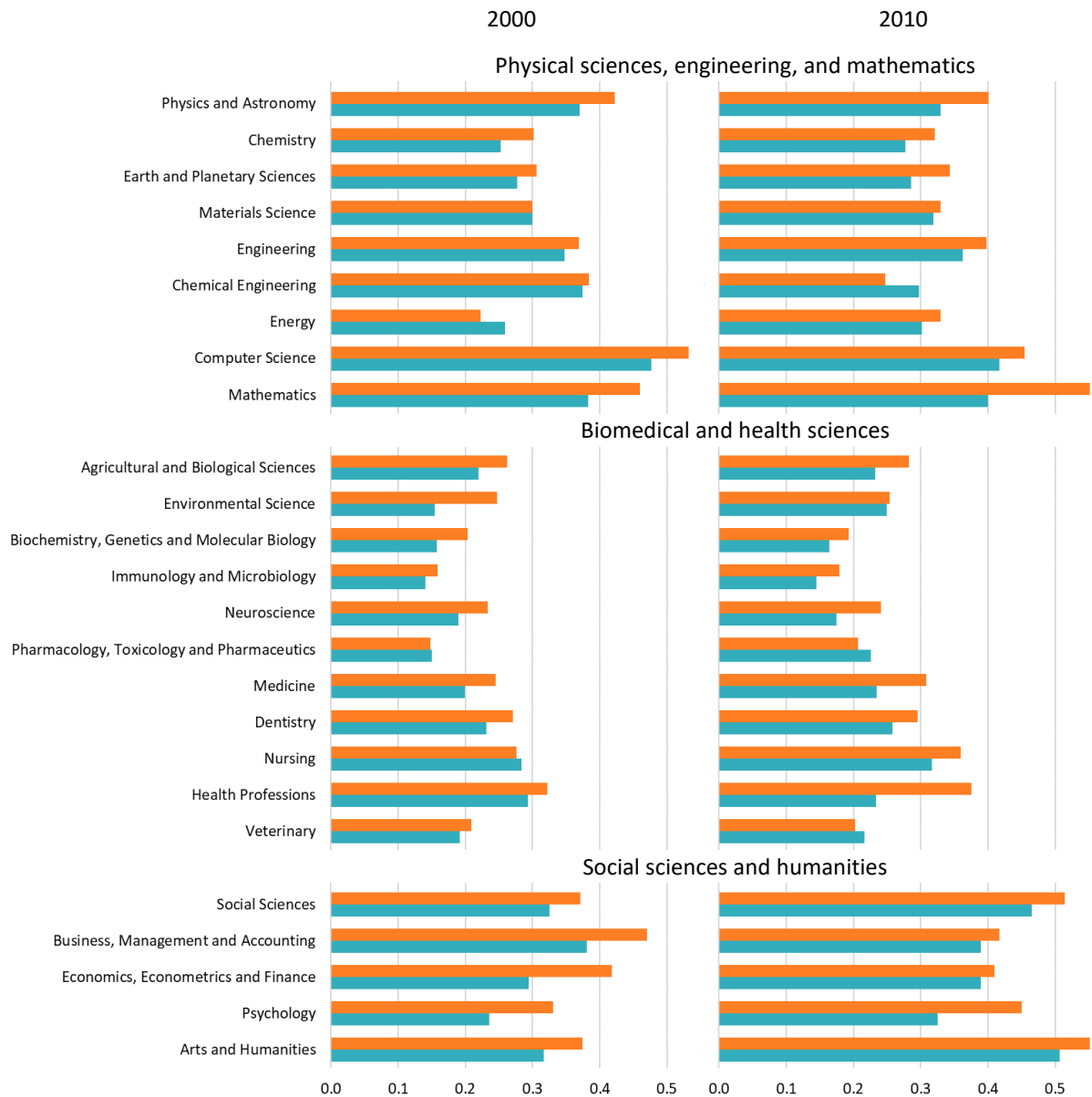

Figure 8. Average number of publications of men and women in year 6 of their career as publishing researcher. Disciplinary statistics for researchers that started their publication career in 2000 or 2010 based on a fractional counting approach.

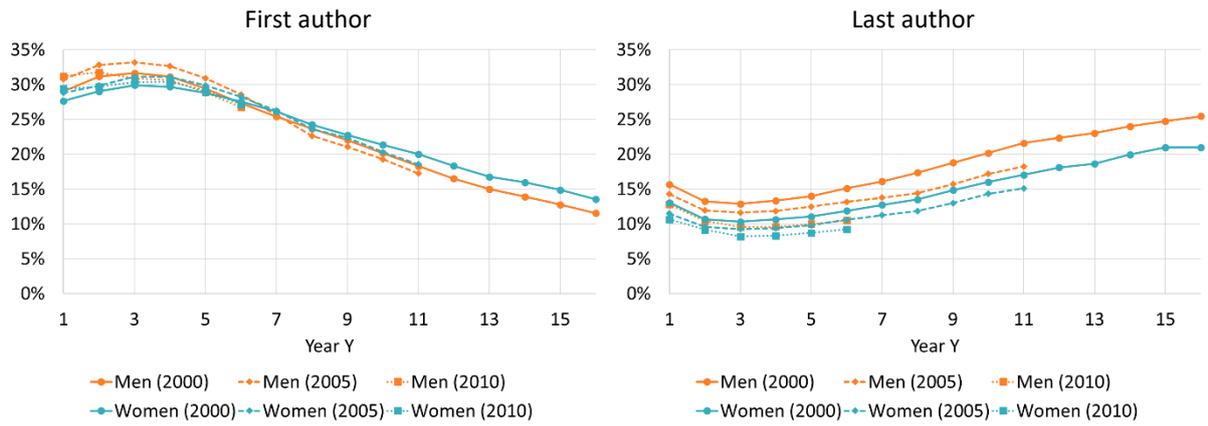

Figure 9. Probability of being first (left) or last (right) author of a publication for men and women in year Y of their career as publishing researcher. Statistics for researchers that started their publication career in 2000, 2005, or 2010.

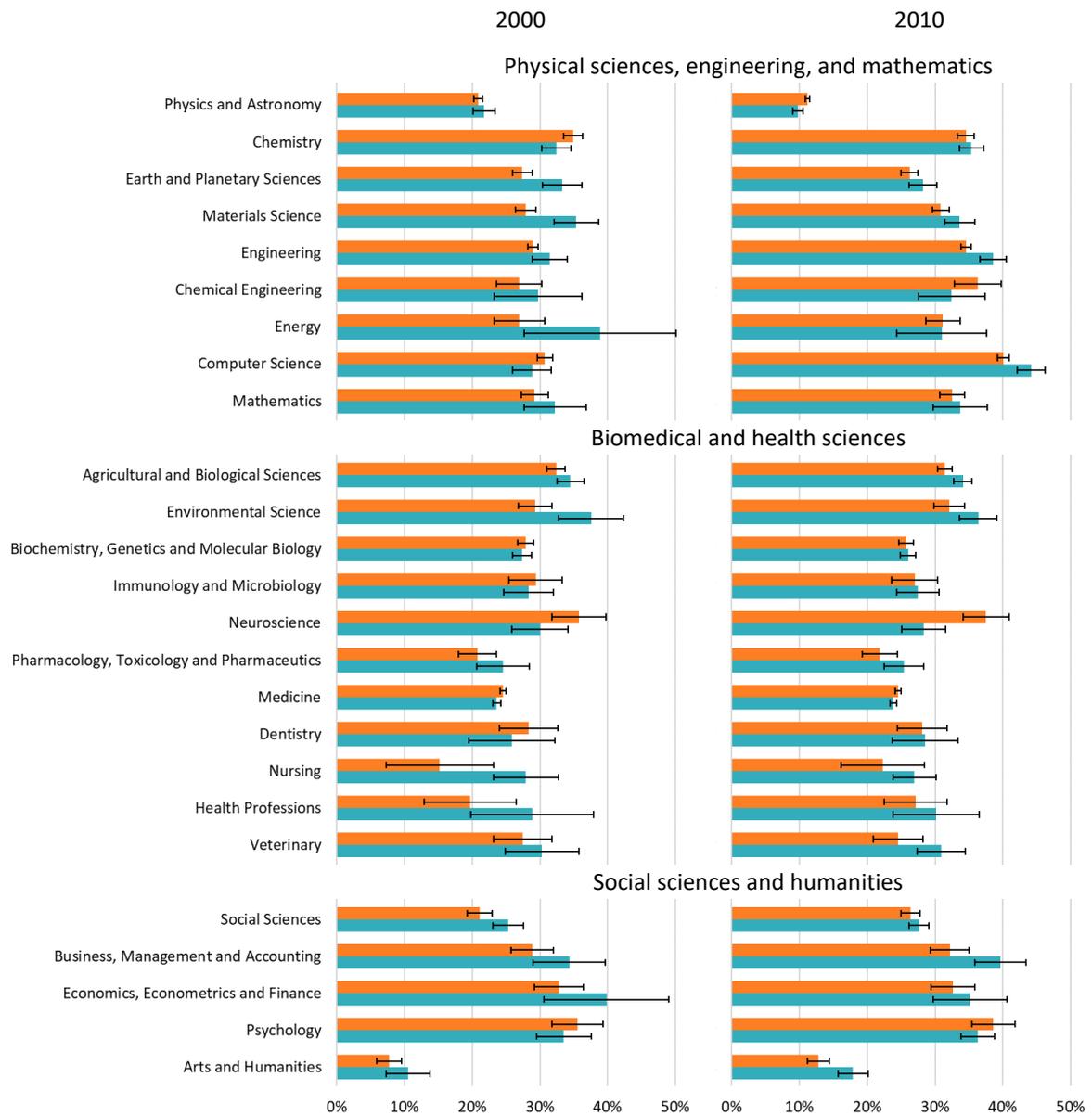

Figure 10. Probability of being first author of a publication for men and women in year 6 of their career as publishing researcher. Disciplinary statistics for researchers that started their publication career in 2000 or 2010. Error bars represent 95% confidence intervals.

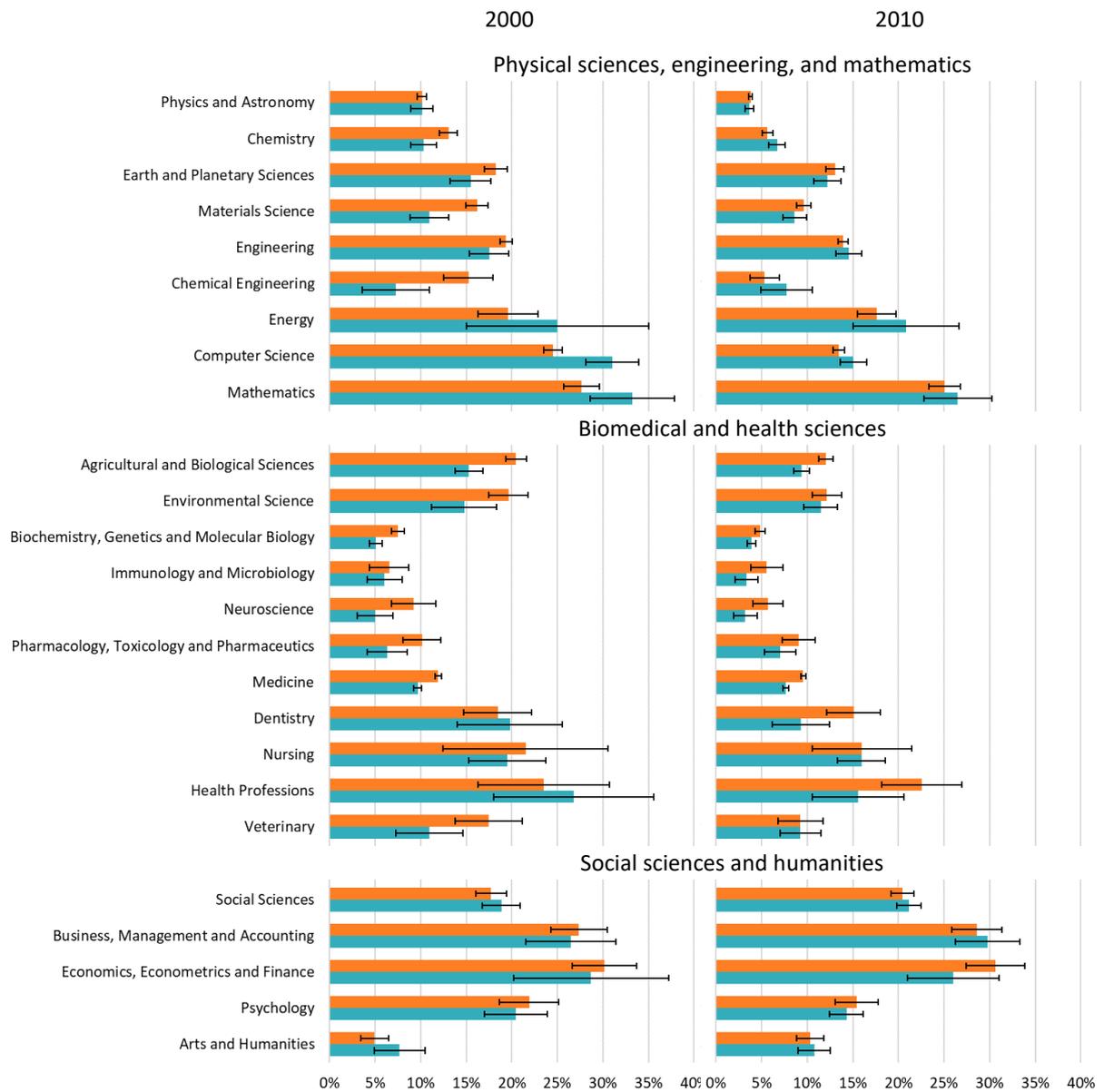

Figure 11. Probability of being last author of a publication for men and women in year 6 of their career as publishing researcher. Disciplinary statistics for researchers that started their publication career in 2000 or 2010. Error bars represent 95% confidence intervals.

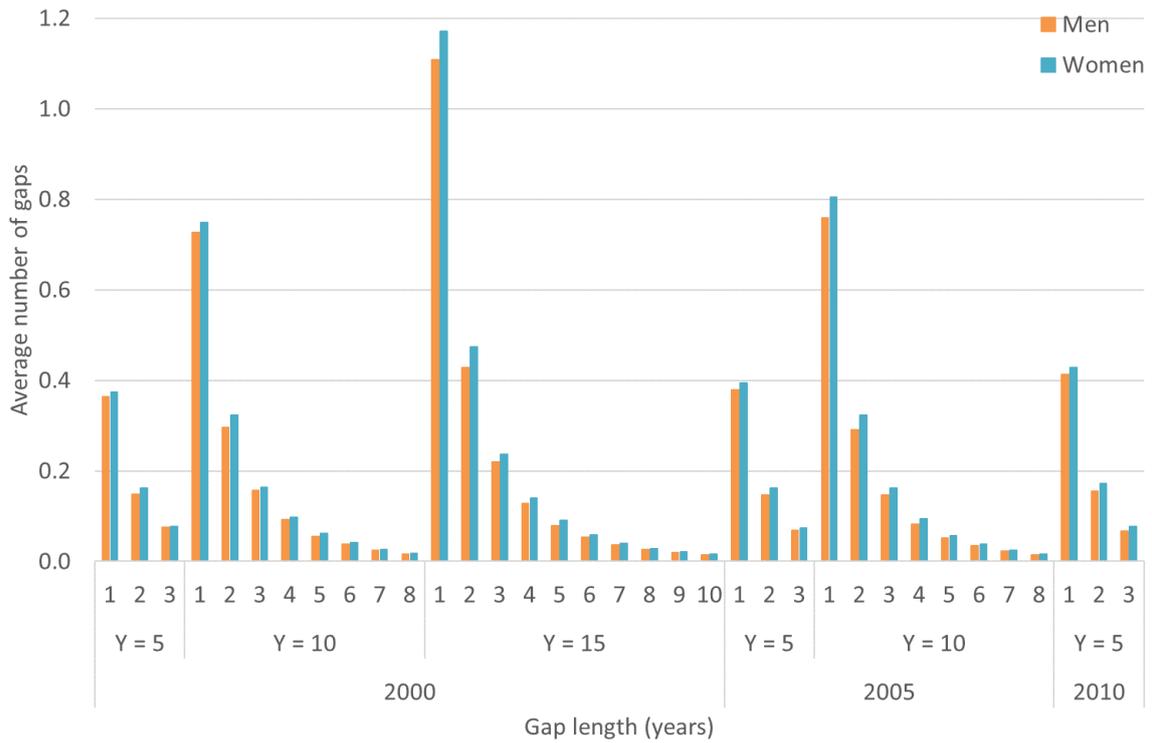

Figure B1. Average number of gaps of a given length in the publication careers of male and female researchers. Statistics for researchers that started their publication career in 2000, 2005, or 2010 and that had a career of at least Y years, where Y equals 5, 10, or 15. For instance, for male researchers that started their publication career in 2005 and that had a career of at least 10 years, the average number of two-year gaps in the first 10 years of their career equals 0.29. A two-year gap is a period of two consecutive years in which a researcher produced no publications.